\documentclass[conference]{IEEEtran}
\IEEEoverridecommandlockouts
\usepackage{cite}
\usepackage{amsmath,amssymb,amsfonts}
\usepackage{graphicx}
\usepackage{textcomp}
\usepackage{xcolor}
\usepackage{comment}
\usepackage{algorithm}
\usepackage{algpseudocode}
\usepackage{romannum}

\def\BibTeX{{\rm B\kern-.05em{\sc i\kern-.025em b}\kern-.08em
    T\kern-.1667em\lower.7ex\hbox{E}\kern-.125emX}}

\begin{document}

\title{GDOP Based BS Selection for Positioning in mmWave 5G NR Networks
}

\author{A. Indika Perera, K. B. Shashika Manosha, Nandana Rajatheva, and Matti Latva-aho
        \\Centre for Wireless Communications, University of Oulu, Finland.
        \\\{indika.perera, nandana.rajatheva, matti.latva-aho\}@oulu.fi, manoshadt@gmail.com}

\maketitle

 \begin{abstract}

The fifth-generation (5G) of mobile communication supported by millimetre-wave (mmWave) technology and higher base station (BS) densification facilitate to enhance user equipment (UE) positioning.
Therefore, 5G cellular system is designed with many positioning measurements and special positioning reference signals with a multitude of configurations for a variety of use cases, expecting stringent positioning accuracies. One of the major factors that the accuracy of a particular position estimate depends on is the geometry of the nodes in the system, which could be measured with the geometric dilution of precision (GDOP). Hence in this paper, we investigate the time difference of arrival (TDOA) measurements based UE positioning accuracy improvement, exploiting the geometric distribution of BSs in mixed LOS and NLOS environment. We propose a BS selection algorithm for UE positioning based on the GDOP of the BSs participating in the positioning process. Simulations are conducted for indoor and outdoor scenarios that use antenna arrays with beam-based mmWave NR communication. Results demonstrate that the proposed BS selection can achieve higher positioning accuracy with fewer radio resources compared to the other BS selection methods.

 \end{abstract}


 \section{Introduction}

The use of a radio signal for positioning and navigation has a long history. Among them as one of the most widely used wireless systems, cellular systems has supported positioning since the first generation (1G).  Even though they are originally designed for communication purposes every generation of cellular systems from 2G onwards were supported user equipment (UE) positioning. All the cellular positioning technologies make use of signal measurements from cellular base stations (BS) and devices, therefore, typically use the existing cellular infrastructure whether they are designed specifically for that purpose or not. Hence, there are many standard defined as well as undefined measurements and methods that are in use for cellular positioning.

Initially, the introduction of the FCC E911 requirements encouraged the study of accurate localization in cellular systems but as positioning accuracy and capabilities increase other commercial use cases were also started using cellular positioning \cite{DelPeral-Rosado2018}. Accuracy requirements for such use cases are broad and include many versatile use cases such as wearables, industry, automotive, self-driving, drones, logistic and tracking. Although global navigation satellite systems (GNSS), which originally have been deployed for military purposes, are now widely applied for commercial applications, can meet some of the accuracy requirements of the use cases mentioned above still it suffers from limited coverage in dense urban areas and particularly in indoor environments \cite{IRACON}.
Hence, most of the natural use cases of positioning occur in an indoor environment such as in a house/office
where GNSS positioning availability is limited as opposed to the cellular positioning. 
The long range radio (LoRa), Wi-Fi, Bluetooth, and other wireless networks also have positioning capability with the advantages of low cost, low power consumption and low complexity\cite{Liu2007}. However, their limited bandwidth, power consumption, and complexity result in a low positioning accuracy as well.

The fifth generation (5G) of mobile communication mainly supported by millimetre-wave (mmWave) multiple-input-multiple-output (MIMO) technology where both UE and BS are equipped with antenna arrays with a large number of antennas. Further, it will operate in high carrier frequencies probably beyond 24 GHz providing large bandwidth for communication thus, high data rates. These favourable properties along with higher network densification, beamforming with narrow beams and precise angle estimation with multiple antennas pave the path of achieving precise positioning with 5G  than the earlier generations of cellular networks. Thus, the 3rd generation partnership project (3GPP) puts higher requirements on the positioning accuracy of the 5G system during 5G NR release 16. New NR positioning reference signals (PRS) for downlink (DL) and sounding reference signal (SRS) for uplink (UL), round-trip time (RTT) measurements with multiple base stations (Multi-RTT),  DL and UL time difference of arrival (TDOA) measurements and BS angle of arrival (AoA) or angle of departure (AoD) measurement can be mentioned as key physical layer technologies that support to achieve user positioning in the 5G system.

The accuracy of a particular position estimate depends on several factors, including the radio ranging measurement accuracy, the algorithm used to process the measurements, and the geometry of the nodes in the system.
Identifying the line-of-sight (LOS)/ non-line-of-sight (NLOS) signals and selecting LOS BSs to estimate position has become a major
approach to mitigate the effect of NLOS caused measurement errors on positioning results.

The geometric dilution of precision (GDOP) is defined as the ratio between the accuracy of a position estimate to the statistical accuracy of the ranging measurements  \cite{Lee1975}. For a terrestrial system, if the location and number of base stations in the desired coverage area are not carefully planned, the GDOP effect can become the dominant factor in limiting the performance of a system. When the angular positions of BSs are close together, the GDOP value is high, resulting in poor positioning performance. For good positional accuracy, the angular position of the transmitting BSs should be such that the receiving UE is “surrounded” by BSs. 

GDOP is a well-investigated metric for the design of GNSS and satellite selection for positioning calculation. Even though GDOP based selection is used for other positioning systems such as satellites\cite{5290368} and ultrasound\cite{8924768}, there is limited literature on using it for cellular user positioning in 5G.
In \cite{Sharp2009} GDOP analysis has been presented for three types of BSs setup where BS are on the circle and where the mobile device is also on the circle, on radials and near a base station. Position accuracy and GDOP analysis are presented in \cite{Sharp2012} for indoor mesh positioning systems in multipath and NLOS propagation environments. Authors in \cite{Deng2020}  propose a GDOP-assisted BS selection method for the hybrid TDOA, RTT and direction of arrival (DOA) positioning in mixed LOS and NLOS indoor open office (IOO) environment where BS geometry is fixed.
As per our knowledge existing literature lacks the knowledge on GDOP based BS selection for mmWave beam-based network especially when the BS geometry is not fixed. Since the positions of the BSs, as well as the number of LOS BSs participating in the positioning calculations, play a major role in achieving high accuracy it is worthwhile to investigate this area using the mmWave 5G NR network for achieving stringent positioning accuracies required by the upcoming generations of cellular networks.

In this paper, we investigate a UE positioning accuracy improvement exploiting the geometric distribution of BSs and the LOS condition of the BS. We present a BS selection criteria for UE position calculation in a mmWave 5G NR network that uses beam based communication between the UE and the randomly distributed BSs. Further, derivation of the GDOP for TDOA based positioning measurements is presented and the proposed BS selection algorithm is based on the calculated GDOP of the BSs in a mixed LOS and NLOS environment. Simulation results for indoor and outdoor scenarios demonstrate that the proposed BS selection can achieve higher positioning accuracy with fewer radio resources.

The rest of the paper is organized as follows. In Section \ref{s2}, we introduce the 5G NR PRS structure, system setup and mathematical derivation of GDOP calculation. Further, in Section \ref{s3} we describe the proposed BS selection algorithm in detail. In Section \ref{s4} we evaluate the performance of the algorithm through simulation results. Finally, in Section \ref{s5}, we summarize our major findings.

\section{System Model}
\label{s2}

\subsection{NR DL positioning reference signals (DL PRS)}

DL PRS resource corresponds to a collection of resource elements arranged in a particular time/frequency pattern where inside each resource element pseudo-random QPSK sequences are transmitted. 
Within a slot, a DL PRS resource can be configured to span 2, 4, 6, or 12 consecutive orthogonal frequency-division multiplexing (OFDM) symbols.  When considering the frequency domain pattern, a DL PRS resource has a comb-like pattern, which means that a QPSK symbol is transmitted on every N-th subcarrier, where N can take the values 2, 4, 6, or 12 \cite{TS38211}. 
The minimum transmission bandwidth of PRS is 24 contiguous physical resource blocks (PRBs) and the maximum transmission bandwidth is 272 PRBs \cite{TS38214}.
 
\subsection{Communication system setup}
\label{comSysSetup}

We consider a MIMO OFDM system in which every BS is equipped with a uniform rectangular array (URA) of $N_t$ antennas and a UE equipped with a URA of $N_r$ antennas operating at a carrier frequency $f_c$ and bandwidth $B$.

We consider BSs transmitting NR PRS to a single UE. At a BS, the NR PRS signal which is generated according to the physical layer cell identity number of the respective BS, is transmitted with $M_t$ number of transmission beams corresponding to each BS beam sweeping direction. At each BS sweep direction, the PRS signal is transmitted again $M_r$ times to corresponds to UE beam sweeping directions. Therefore, there will be a total of  $M_t \times M_r$ number of NR PRS transmissions between a specific BS and the UE.

mmWave scattering MIMO channel which simulates a multipath propagation channel between the BS transmitting array and the UE receiving array is used. Radiated PRSs from a BS transmitting array are reflected from multiple scatterers in the environment before receive to the UE receiving array. This channel generates a BS to UE distance dependent channel delay,  gain, phase change, and atmospheric loss.

\begin{figure}[htbp]
\centerline{\includegraphics[trim={0cm 1cm 0cm 0.5cm},clip,width=0.4\textwidth,keepaspectratio]{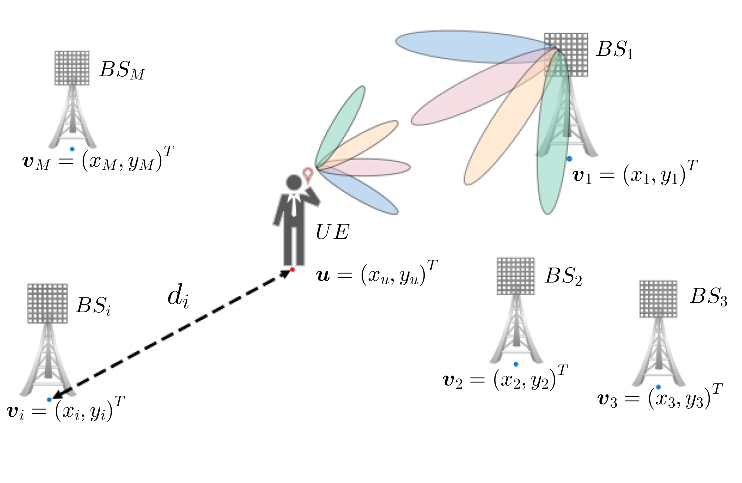}}
\caption{System setup comprising of multiple MIMO BSs and a single UE.}
\vspace{-2mm}
\label{SystemModel}
\end{figure}

\subsection{Positioning system setup}
\label{sysSetup}

We consider that the UE could receive downlink PRSs from $M$ BSs and obtain $M-1$ number of TDOAs with respect to the reference BS. Without loss of generality, we select the nearest BS as the reference BS and denote it as BS 1. The remaining BSs are sorted by the distance to UE and the farthest is denoted $M$.

In the following description, we consider two-dimension Cartesian coordinate system. Hence, the actual position of the UE is denoted by $\boldsymbol u={(x_u,y_u)}^T$ and BS positions are denoted by $\boldsymbol {\boldsymbol v}_i={(x_i,y_i)}^T,\;\;i=1,\;2,\;3,\dots,\;M$ as shown in Fig. \ref{SystemModel}.
The actual distance between $i$th BS and the UE is given by
\vspace{-2mm}
\begin{equation}
    \begin{split}
    d_i=\Arrowvert\boldsymbol {v}_i-\boldsymbol u\Arrowvert =\sqrt{{(x_i-x_u)}^2+{(y_i-y_u)}^2},\\ i=1,2,3,\dots,M.
    \end{split}
\label{eq1}
\end{equation}
The ranging difference between the $i$th and the reference BS, calculated using TDOA measurements can be expressed as follows
\vspace{-4mm}
\begin{equation}
\begin{split}
    \begin{array}{c}
    r_{i,1}=c{(t_i-t_1)}, \;i=2,3,\dots,M,
    \end{array}
\end{split} 
\label{eq2}
\end{equation}
where $r_{i,1}$ is the ranging difference calculated by TDOA measurement, $t_i$ is measured PRS arrival time from $i$th BS to UE, $t_1$ is measured PRS arrival time from BS $1$ to UE and $c$ is the speed of light. Alternatively, the ranging difference can be expressed as follows
\vspace{-2mm}
\begin{equation}
\begin{split}
    \begin{array}{c}
    r_{i,1}=d_i - d_1 + e_{i,1} = d_{i,1} + e_{i,1},\; i=2,3,\dots,M,
    \end{array}
\end{split} 
\label{eq3}
\end{equation}
where $d_{i,1}$ is the actual distance difference between the $i$th and the reference BS and $e_{i,1}$ is the measurement error. Two dimensional TDOA measurement model in matrix form when UE is at an arbitrary location $\mathbf{x} =[x,y]^T$  is
\begin{equation}
    \mathbf{r}= \mathbf{f(x)} + \mathbf{\vec e},
\label{eq4}
\end{equation}
where $\mathbf {r}= [r_{2,1}\; r_{3,1}\dots r_{M,1}]^T,\;\mathbf{\vec e}= [e_{2,1}\;e_{3,1}\dots e_{M,1}]^T$ and 
\vspace{-2mm}
\begin{multline}
\mathbf {f(x)} = \begin{bmatrix}
                  d_{2,1}(\mathbf x) & d_{3,1}(\mathbf x) & \dots & d_{M,1}(\mathbf x)
                   \end{bmatrix} ^T \\ =
               \begin{scriptsize}   \begin{bmatrix}
                  \sqrt{{(x_2-x)}^2+{(y_2-y)}^2}-\sqrt{{(x_1-x)}^2+{(y_1-y)}^2} \\
                  \sqrt{{(x_3-x)}^2+{(y_3-y)}^2}-\sqrt{{(x_1-x)}^2+{(y_1-y)}^2} \\
                  \vdots \\
                  \sqrt{{(x_M-x)}^2+{(y_M-y)}^2}-\sqrt{{(x_1-x)}^2+{(y_1-y)}^2} \\
                 \end{bmatrix}.\end{scriptsize}
\label{vectR}
\end{multline}

\vspace{0.5mm}
\subsection{Localization algorithm}
\label{locAlgo}

In order to find the position of the UE using the TDOA measurements expressed in \eqref{vectR} we construct the least square cost function as

\vspace{-3mm}
\begin{scriptsize}
\begin{align}
\mathbf {J(x)} =& \sum_{i=2}^{M}  \left (r_{i,1} -\sqrt{{(x_i-x)}^2+{(y_i-y)}^2}+ \sqrt{{(x_1-x)}^2+{(y_1-y)}^2} \right)^2 ,\notag
\end{align} 
\end{scriptsize}in matrix form
\vspace{-2mm}
\begin{equation}
\mathbf {J(x)} = (\mathbf{r}-\mathbf{f(x)})^T (\mathbf{r}-\mathbf{f(x)}),
\label{costFunVec}
\end{equation}
and the least square position estimate is
\vspace{-1mm}
\begin{equation}
\mathbf {\hat x } = \arg \min _ x \mathbf {J(x)}.
\label{argMIn}
\end{equation}
\vspace{-4mm}

The minimum value of the \eqref{argMIn} is achieved via the steepest descent iterative procedure where the iterative step is calculated using 
\vspace{-2mm}
\begin{equation}
\mathbf {\hat x_{k+1}} = \mathbf {\hat x_{k} } -\mu \nabla \mathbf {J(\hat x)},
\label{sd}
\end{equation}
where $\mu$ is a positive constant, which controls the convergence rate and stability, $\nabla \mathbf {J(\hat x)}$ is the gradient vector computed at the $k$th iteration estimate.

\subsection{GDOP for the TDOA based positioning}

Let  $(x',y')$ be the estimated position and  $ g_{i,1}(x',y'),\; i=2,3,\dots,M$ be the functional relationships of estimated position and TDOA measurement errors. For a system containing $M$ BSs, the TDOA measurement error equations can be written as
\vspace{-2mm}
\begin{equation}
\begin{split}
   \begin{array}{c}
   e_i=g_{i,1}{(x',y')}=r_{i,1}-\sqrt{{(x_i-x')}^2+{(y_i-y')}^2} \\+\sqrt{{(x_1-x')}^2+{(y_1-y')}^2},\;i=2,3,\dots,M.
   \end{array} 
\label{eq8}
\end{split}
\end{equation}

In order to calculate the GDOP for TDOA positioning, we linearize the measurement equations using the first order Taylor series. We can approximate $ g_{i,1}(x',y'),\; i=2,3,\dots,M$ when UE locates at $(x_\circ, y_\circ)$, as
\vspace{-2mm}
\begin{equation}
\begin{split}
    g_{i,1}{(x',y')}\approx g_{i,1}{(x_\circ,y_\circ)}+{(x'-x_\circ)}\frac{\partial g_{i,1}{(x_\circ,y_\circ)}}{\partial x} \\+{(y'-y_\circ)}\frac{\partial g_{i,1}{(x_\circ,y_\circ)}}{\partial y} ,\; i=2,3,\dots,M.
\label{eq9}
\end{split}
\end{equation}

When there are no TDOA measurement errors, the estimated position $(x',y')$ and the UE’s actual location $ (x_u,y_u)$ would be same, thus, \eqref{eq9} is rewritten as
\vspace{-2mm}
\begin{equation}
\begin{split}
    0=g_{i,1}{(x_\circ, y_\circ)}+{(x_u-x_\circ)}\frac{\partial g_{i,1}{(x_\circ, y_\circ)}}{\partial x} \\ +{(y_u-y_\circ)}\frac{\partial g_{i,1}{(x_\circ, y_\circ)}}{\partial y},\; i=2,3,\dots,M.
\label{eq10}
\end{split}
\end{equation}

When there are TDOA measurement errors, using $e_i,\; i=2,3,\dots,M$ to express the TDOA measurement errors \eqref{eq9} is rewritten as
\vspace{-2mm}
\begin{equation}
\begin{split}
    e_i=g_{i,1}{(x_\circ, y_\circ)}+{(x'-x_\circ)}\frac{\partial g_{i,1}{(x_\circ, y_\circ)}}{\partial x} \\ +{(y'-y_\circ)}\frac{\partial g_{i,1}{(x_\circ, y_\circ)}}{\partial y},\; i=2,3,\dots,M.
\label{eq11}
\end{split}
\end{equation}

Let position estimation error vector be $\triangle \mathbf u={[e_x,e_y]}^T$, where $e_x=x'-x_u$, $e_y=y'-y_u$. Subtracting \eqref{eq10} from \eqref{eq11}, the relationship between $\triangle\mathbf u$ and TDOA measurement errors $e_i, \; i=2,3,\dots,M$ can be presented as
\begin{align}
 e_i=&e_x\frac{\partial g_{i,1}{(x_\circ,y_\circ)}}{\partial x}+e_y\frac{\partial g_{i,1}{(x_\circ,y_\circ)}}{\partial y}
 \end{align}
\vspace{-4mm}
\begin{scriptsize}
\begin{align}
 =& e_x{(\frac{x_i-x_\circ}{\sqrt{{(x_i-x_\circ)}^2+{(y_i-y_\circ)}^2}}-\frac{x_1-x_\circ}{\sqrt{{(x_1-x_\circ)}^2+{(y_1-y_\circ)}^2}})}\notag\\+&e_y{(\frac{y_i-y_\circ}{\sqrt{{(x_i-x_\circ)}^2+{(y_i-y_\circ)}^2}}-\frac{y_1-y_\circ}{\sqrt{{(x_1-x_\circ)}^2+{(y_1-y_\circ)}^2}})} \notag.
\label{eq12} 
\end{align}
\end{scriptsize}

After linearization,TDOA measurement equations in \eqref{eq8} can be presented as the following matrix form:
\vspace{-2mm}
\begin{equation}
   {\mathbf e} = \mathbf A\triangle\mathbf u,
\label{eq13}
\end{equation}where
\vspace{-2mm}
\begin{equation}
\begin{split}
\mathbf e=\begin{bmatrix}
                    e_2 \\
                    e_3 \\
                    \vdots \\
                    e_M
                    \end{bmatrix}, 
\mathbf A = \begin{bmatrix}
    \alpha_2 &\beta_2 \\
    \alpha_3 &\beta_3 \\
    \vdots   & \vdots \\
    \alpha_M & \beta_M
    \end{bmatrix},
\alpha_i=\frac{\partial g_{i,1}{(x_\circ,y_\circ)}}{\partial x},\;\\\beta_i=\frac{\partial g_{i,1}{(x_\circ,y_\circ)}}{\partial y},\; i=2,3,\dots,M.
\end{split}
\notag
\end{equation}

For different TDOA measurements, the measurement errors $e_{i}, i=2,\dots,M$ can assumed to be independent and identically distributed. Thus, for all elements in the error vector $\mathbf e$,
\vspace{-6mm}
\begin{align}
   \mathbb E{(e_i)}= & 0,i=2,3,\dots,M, \label{eqMean}\\
    Cov{(e_ie_j)}= & 0,i\neq j\; \text{and}\;i,\;j=2,3,\dots,M.
\end{align}Assume that standard deviations of the TDOA measurement errors are equal. Hence,
\vspace{-1mm}
\begin{equation}
Var{(e_i)}=\sigma_{tdoa}^2,i=2,3,\dots,M,
\label{eq16}
\end{equation} where $\sigma_{tdoa}^2$ is the variance of TDOA measurement errors. The weighted least square (WLS) solution to \eqref{eq13} is given by
\vspace{-1mm}
\begin{equation}
    \triangle\mathbf u=\mathbf{{(A^TWA)}^{-1}A^TW}\mathbf e,
\label{eq17}    
\end{equation} where $ \mathbf W$ is the weighted matrix, which can be obtained by calculating the covariance of the TDOA measurement errors as
\vspace{-2mm}
\begin{equation}
\begin{split}
Cov({\mathbf e}) = \mathbb E \begin{pmatrix}
                                            \begin{bmatrix}
                                            e_2 \\
                                            e_3 \\
                                          \vdots \\
                                            e_M
                                            \end{bmatrix} 
                                            \begin{bmatrix}
                                            e_2 & e_3 & \dots & e_M
                                            \end{bmatrix}
                                    \end{pmatrix} \\
                                    = \sigma_{tdoa}^2 \begin{bmatrix}
                                             1 & 0 &   \dots & 0 \\ 
                                             0 & 1 &   \dots & 0 \\ 
                                        \vdots & \vdots  & \ddots & \vdots \\ 
                                             0 & 0 &  \dots & 1 \\ 
                                            \end{bmatrix} = \sigma_{tdoa}^2 \mathbf \Sigma.
\end{split}
\label{eq18} 
\end{equation}
\begin{equation}
  \mathbf W= \mathbf \Sigma^{-1}   \label{eq19} 
\end{equation} Using \eqref{eqMean}, the mean of $\triangle\mathbf u $ can be calculated as
\begin{align}
    \mathbb E{(\triangle \mathbf u)}=& \mathbb E{\left [\mathbf{{(A^TWA)}^{-1}A^TW}{\mathbf e}\right ]} \notag \\ =& \mathbf {{(A^TWA)}^{-1}A^TW }\mathbb E{({\mathbf e})}=0. \label{eq20} 
\end{align} The covariance matrix of estimated position error $\triangle\mathbf u $  can be expressed as
\vspace{-2mm}
\begin{align}
    Cov{(\triangle\mathbf u)}=&\mathbb  E{[\triangle\mathbf u\triangle\mathbf u^T]}= \mathbb E{ {[(\mathbf A^T{({\mathbf e}{\mathbf e}^T)}^{-1}\mathbf A)}^{-1}]} \notag \\ =& {( \mathbf A^TCov{({\mathbf  e})}^{-1}\mathbf A)}^{-1}.
\label{eq21}  
\end{align} Further, using the results in \eqref{eq18} and \eqref{eq19} the covariance matrix in \eqref{eq21} is calculated as
\vspace{-2mm}
\begin{align}
    Cov{(\triangle\mathbf u)}= {(\mathbf A^T\frac1{\sigma_{tdoa}^2}\mathbf {WA})}^{-1} = \sigma_{tdoa}^2{(\mathbf {A}^T\mathbf {WA})}^{-1}.
\label{eq22}  
\end{align}

 GDOP is defined as the ratio of positioning error to positioning signal measurement error \cite{Lee1975}.
The GDOP which is calculated using WLS is called weighted GDOP \cite{Chen2013} and the GDOP of the TDOA based positioning is given by
\vspace{-1mm}
\begin{align}
    \mathrm{GDOP}=&\sqrt{\frac{Var{(e_x)}+Var{(e_y)}}{\sigma_{tdoa}^2}} \notag   \\=&\sqrt{\frac{tr{(Cov{(\triangle \mathbf u)})}}{\sigma_{tdoa}^2}} =\sqrt{tr{({\mathbf {(A^TWA)}}^{-1})}}.\label{eq23}  
\end{align}

 \section{Methodology}
\label{s3}

In this section, the NR PRS beam sweeping, measurement procedure and the proposed BS selection algorithm for UE positioning based on the GDOP of the BSs participating in the positioning process are presented in detail.

 \subsection{NR PRS beam sweeping and measurements}
Following the typical  NR positioning procedure, location management function (LMF) in the 5G core network informs the BSs and the UE to start the positioning process by providing required assistant data. These assistant data required for DL-TDOA contains a list of candidate BSs and their positions together with their DL-PRS configurations. Multiple DL-PRS resources are configured to be transmitted from each candidate BS corresponds to a different transmit beam of the BSs. We assume that BSs in the network are time-synchronized with each other via the backhaul network and also all the participating BSs are transmitting the same number of beams. Hence, the BSs participating in the process, transmit PRSs in a synchronized manner via beam sweeping. UE receives these PRS transmissions with beam sweeping with the number of receive beams configured in UE. Therefore, the PRS transmission repetition factor in the DL-PRS configurations is selected based on the number of UE beams.

Once, all the PRS transmissions are performed, UE calculates the received power and time of arrival (ToA) of the PRS for each of the UE receive beams and transmit beams of every candidate BSs that are participated in the localization process. Then the UE selects transmit and receive beam pair with the highest received power from each BSs as the best beam pair for communication between that BS and UE. UE calculates the ToA of each beam pairs by correlating the received PRS signal with a generated copy of that same signal using the PRS configuration data shared at the beginning of the process. UE selects the minimum ToA value from the set of ToA values that corresponds to all the transmit and receive beam pairs between a BS and UE as the ToA between that specific BS and UE. These ToA values are used to calculate the time difference of arrival with respect to the reference BS that is selected during the BS selection algorithm explain in  \ref{BSselAlgo}.

Assuming channel conditions between all the BSs and UE is known to the UE, UE categorizes the channel between each BS and UE into LOS or NLOS. We denote $BS_{los}$ as the LOS BS set and $BS_{nlos}$ as the NLOS BS set. Further, the number of BSs in each of the sets are denoted as $N_{los}$ and $N_{nlos}$, respectively.

 \subsection{Proposed BS selection algorithm}
\label{BSselAlgo} 

Once, the UE categorized BSs into LOS and NLOS, it uses the BSs in the $BS_{los}$ set to calculate an initial position value using the localization algorithm presented in section \ref{locAlgo}. As the first step of the BS selection method, UE calculates the GDOP of all the BSs in the $BS_{los}$ set at the initial position estimate calculated earlier. Here it calculates $N_{los} $ number of GDOP values corresponds to GDOP when each BS is used as the reference BS. Then these LOS BSs are sorted according to the ascending order of the calculated GDOP values. Similarly, UE calculates the GDOP of the BSs in the $BS_{nlos}$ set and these NLOS BSs are also sorted according to their ascending GDOP values as earlier.

Then the UE selects $N$ number of BSs from the participating BSs to calculate the position of the UE. At the selection, UE gives priority to the GDOP sorted $BS_{los}$ set first and then the remaining number of BS required to satisfy the $N$ number of BSs are selected from the GDOP sorted $BS_{nlso}$ set.  Hence, the LOS BS with the lowest GDOP is the first BS in the selected set and other LOS BSs follow it with increasing GDOP and at last, the NLOS BSs sorted according to their GDOP are in the selection. The first BS which is the LOS BS with the lowest GDOP is selected as the reference BS for calculating the TDOA values needed for the UE position calculation. Finally, the position of the UE is calculated using the selected $N$ number BSs including the reference BS from the above selection criteria and with the least square position estimate presented earlier.

 \section{Simulations and Results }
\label{s4}

\begin{figure}[htbp]
\centerline{\includegraphics[trim={0cm 0.2cm 1.0cm 0.5cm},clip,width=0.45\textwidth,keepaspectratio]{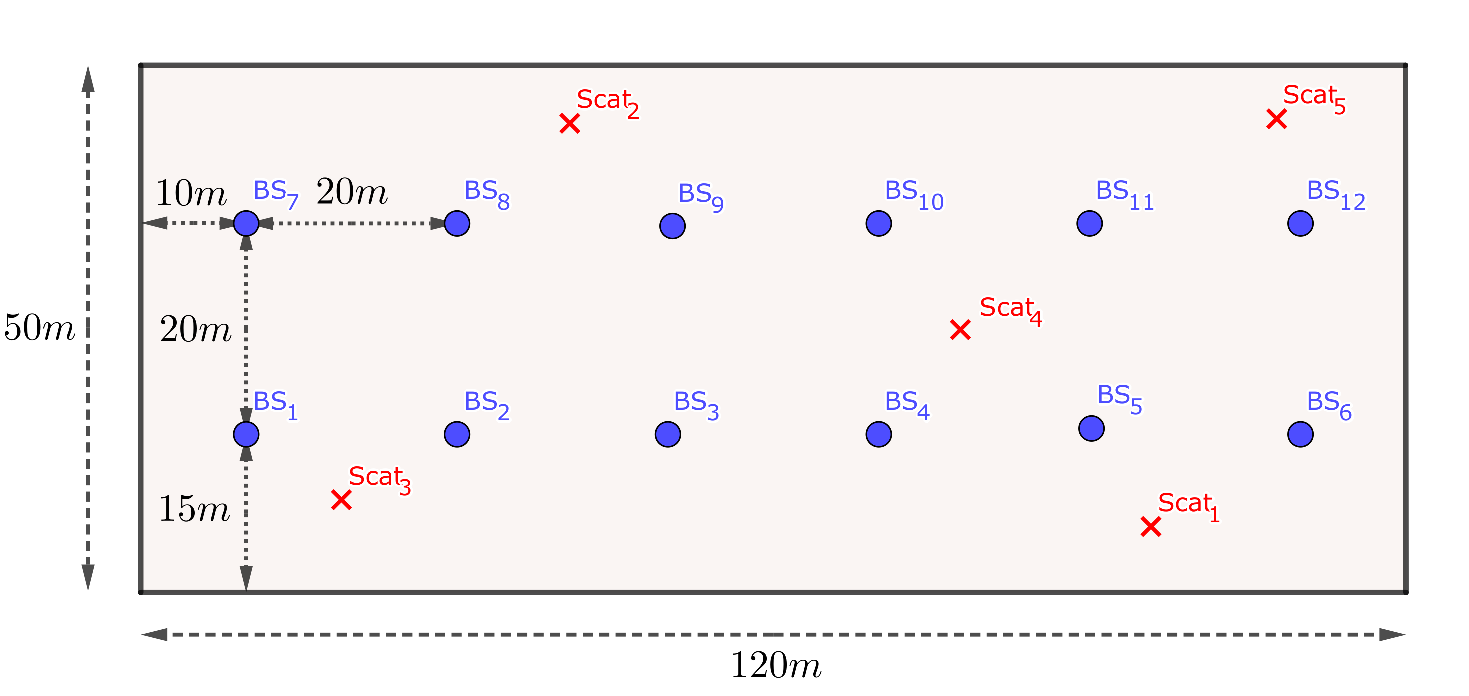}}
\caption{Fixed BS locations of the indoor open office scenario with some scatterers located randomly.}
\vspace{-4mm}
\label{12BSIOO}
\end{figure}

We simulate the proposed BS selection technique using the system setup mentioned in section \ref{s2}. We have simulated an urban microcell (UMi) scenario with 7 BSs located at random positions and an IOO scenario that includes 12 BSs with fixed positions as shown in Fig. \ref{12BSIOO}. In both scenarios, we simulate a mixed LOS and NLOS environment where UE has a LOS condition only with some randomly selected BSs. Main simulation parameters are listed in Table \ref{tab1} below.
\vspace{-4mm}
\begin{table}[htbp]
\caption{Simulation Parameters}
\begin{center}
\begin{tabular}{|l|c|}
\hline
\textbf{Description}&{\textbf{Value}} \\
\hline
Carrier frequency & 28 GHz \\
\hline
Sub carrier spacing & 120 kHz \\
\hline
Number of PRBs & 56 \\
\hline
Array type and size (Tx and Rx)  & URA  8x8 and 4x4 \\
\hline
Number of beams (Tx and Rx)  & 16 and 8 \\
\hline
PRS comb size & 12 subcarriers\\
\hline
PRS time domain size & 12 symbols \\ 
\hline
\end{tabular}
\label{tab1}
\end{center}
\end{table}

We use 4 BSs for calculating the UE position since a minimum of 4 BSs (including reference BS) are required to achieve a unique position estimate using the TDOA measurement equations. We calculate the position of the UE using 4 BSs selected from our proposed BS selection method and for comparison purposes, we calculate the position of the UE using 4 BSs selected randomly from all available BSs and the closest BS from that selection is considered as the reference BSs for TDOA calculations. Further, we use another common selection method named distance based BS selection where the nearest LOS BSs are selected first and then nearest NLOS BSs are selected for the remaining required number of BSs. In this situation nearest LOS BS would be selected as the reference BS for TDOA calculations. The least-square position estimate is used in position calculations in all the selection methods and simulation scenarios.
\vspace{-2mm}
  
\subsection{Urban micro cell scenario}

\begin{figure}[htbp]
\centerline{\includegraphics[trim={0.5cm 0cm 1cm 0.7cm},clip,width=0.42\textwidth,keepaspectratio]{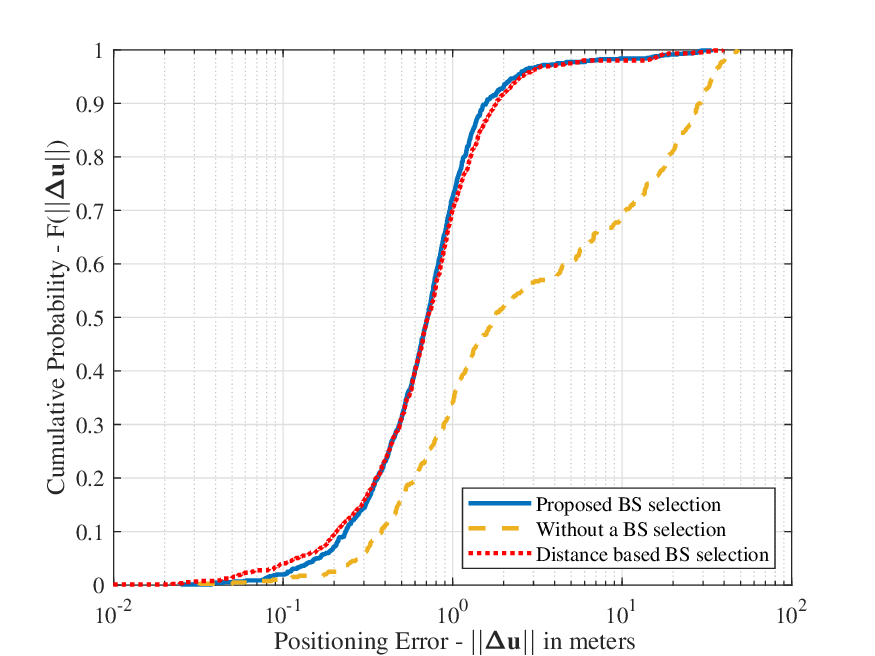}}
\caption{CDF of the positioning error in UMi with different BS selection methods for positioning.}
\vspace{-2mm}
\label{7BSResult}
\end{figure}

For the micro cell environment, the BSs are randomly located maintaining a minimum of 100m inter base station distance in $x$ and $y$ Cartesian axis directions.
We simulate single UE in the UMi environment and the channel between the UE and BSs are simulated as mentioned in section \ref{comSysSetup} with 5 scatters in random locations. This setup is simulated for many BSs and scattering location instances and the user position is estimated for each instance. Then the error of the estimated position with respect to the true location is calculated as $\lVert \triangle \mathbf u \rVert$ for each instance. Fig. \ref{7BSResult} shows the cumulative distribution function (CDF) of the error of the estimated position for all BS selection methods.
\vspace{-2mm}

Results in Fig. \ref{7BSResult} show that the positioning error is reduced when the proposed BS selection algorithm is used for position calculation compared to the random selection of BSs and distance-based BS selection. It could be seen that the proposed BS selection algorithm outperforms the random BS selection methods used for the comparison. Even though the proposed BS selection is comparable with the distance-based BS selection, the proposed BS selection has 90\% of its positioning error values below 1.55m while in the distance-based BS selection corresponding value is 1.80m. It could be seen that these values are on par with most of the positioning error values currently available in the industry \cite{Dwivedi}. We are not capable of exactly reproduce the results in \cite{Dwivedi} for comparison using the available limited resources since those are obtained via system-level simulators. Further, the proposed BS selection method accuracy satisfies the minimum horizontal positioning error targets required for regulatory use cases and commercial use cases as presented in \cite{3GPP2019}. Hence, the proposed BS selection method could be used for higher accuracy requirements where the computational power is not limited while the less complex distance-based method could be used on devices with low computational capability.

\subsection{Indoor open office scenario}

\begin{figure}[htbp]
\centerline{\includegraphics[trim={0.5cm 0cm 1cm 0.7cm},clip,width=0.42\textwidth,keepaspectratio]{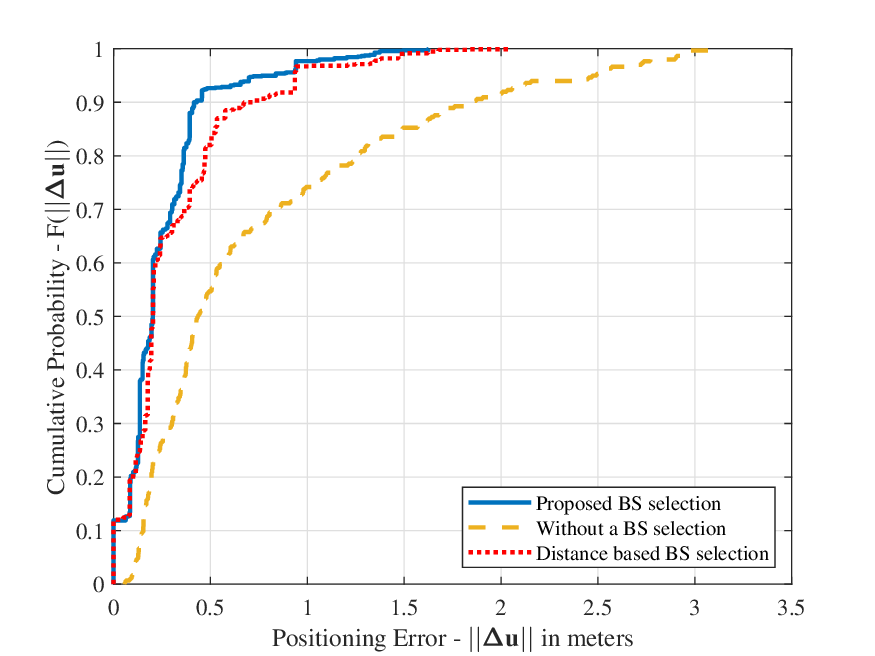}}
\caption{CDF of the positioning error in IOO with different BS selection methods for positioning.}
\vspace{-2mm}
\label{12BSResult}
\end{figure}

Indoor open office scenario is simulated according to the 3GPP TR 38.901 indoor office scenario \cite{3GPP2018}. As shown in Fig. \ref{12BSIOO} it has fixed BS locations and a minimum of 20m inter BS distance in $x$ and $y$ Cartesian axis directions. We simulate a single UE in the IOO environment and the channel between the UE and BSs are simulated as earlier with 5 scatters in random locations. For the calculations of UE position 4 BSs are used and a CDF plot for calculated positioning error is drawn for the proposed BS selection, distance-based selection and the random BS selection as shown in Fig. \ref{12BSResult}.
\vspace{-2mm}

Results in Fig. \ref{12BSResult}  show that the proposed BS selection algorithm provides better positioning capability and low positioning error compared to distance-based BS selection and random BS selection. For the proposed BS selection algorithm 90\% of the positioning error values are below 0.45 m and for the distance-based BS selection algorithm, it is 0.8 m. Further, those satisfy the minimum horizontal positioning error targets required for regulatory and commercial use cases. Compared to the previous scenario the proposed BS selection algorithm provides more accuracy improvement compared to the computational complexity added to the system. Since this scenario is an indoor environment, the positions of the BSs are not dynamically changing as opposed to an outdoor scenario. Therefore, considering the low calculation frequency compared to an outdoor scenario, using the proposed BS selection could provide a higher device efficiency as opposed to using it in an outdoor scenario.

Further, positioning accuracy values we achieve for both scenarios are simulated with less bandwidth (56 PRBs) compared to the existing results where most of them are achieved via simulating the maximum allowable bandwidth (272 PRBs) for PRS signal. Hence, we could state that our proposed algorithm required fewer radio resources in achieving the required positioning accuracies contributing to increase the network efficiency.

 \section{Conclusion}
\label{s5}

In this paper, we investigate a UE positioning accuracy improvement exploiting the geometric distribution of BSs in mixed LOS and NLOS environment. GDOP provides a measure of the geometry of the nodes in a system which is one of the major factors that the accuracy of a particular position estimate depends on. We proposed a BS selection algorithm for UE positioning based on the GDOP of the BSs used for calculating the position. We derive the GDOP calculation method for TDOA based downlink positioning measurements. Simulations are conducted for the mmWave antenna arrays with beam-based communication in indoor and outdoor scenarios and results demonstrate that the proposed BS selection can achieve higher positioning accuracy with fewer radio resources. Classification of LOS and NLOS BS set used for the algorithm using the measured data itself would be a possible future improvement. Further, the algorithm can extend to improve the positioning and tracking accuracy of a moving user.

\vspace{-2mm}
\bibliographystyle{IEEEbib}

\bibliography{main}

\end{document}